\documentclass[aps,prx,twocolumn,showpacs,preprintnumbers,amsmath,amssymb] {revtex4-1}
\usepackage{amssymb}
\usepackage{dcolumn}
\usepackage{color}
\usepackage{graphicx}
\usepackage[pdftex,colorlinks=true,linkcolor=red,citecolor=blue]{hyperref}
\usepackage{soul}

\begin{document}

\title{On the magnetic and electronic properties of NpPdSn}

\author{K.~Shrestha,$^{1,\dagger}$ D.~Antonio,$^{1,\dagger}$ J-C.~Griveau,$^{2}$ K.~Proke\v{s},$^{3}$ P.~Gaczy\'{n}ski,$^{2}$  E.~Colineau,$^{2}$ R.~Caciuffo,$^{2}$ and K. Gofryk$^{1,}$}

\email{Email address: krzysztof.gofryk@inl.gov}

\affiliation{$^{1}$Idaho National Laboratory, Idaho Falls, Idaho, 83415, USA\\$^{2}$DG Joint Research Centre, Directorate G-Nuclear Safety and Security, D-76125 Karlsruhe, Germany\\$^{3}$Helmholtz-Zentrum Berlin, Hahn-Meitner-Platz 1, D-141 09 Berlin, Germany}

\begin{abstract}
We have studied NpPdSn by means of the heat capacity, electrical resistivity, Seebeck and Hall effect, $^{237}$Np M\"{o}ssbauer spectroscopy, and neutron diffraction measurements in the temperature range 2~-~300~K and under magnetic fields up to 14~T. NpPdSn orders antiferromagnetically below the N\'eel temperature $T_N$ = 19~K and shows localized magnetism of Np$^{3+}$ ion with a a doubly degenerate ground state. In the magnetic state the electrical resistivity and heat capacity are characterized by electron-magnon scattering with spin-waves spectrum typical of anisotropic antiferromagnets. An enhanced Sommerfeld coefficient and typical behavior of magnetorestistivity, Seebeck and Hall coefficients are all characteristic of systems with strong electronic correlations. The low temperature antiferromagnetic state of NpPdSn is verified by neutron diffraction and $^{237}$Np M\"{o}ssbauer spectroscopy and possible magnetic structures are discussed.

\end{abstract}

\thanks{These two authors contributed equally}

%
%

%

\maketitle
%
%

\section {Introduction}

In the last decades, study of the physics of actinide-based intermetallics has been stimulated by a large variety of exotic physical phenomena displayed in this class of materials. These interesting behaviors are mainly coming from the hybridization of 5$f$-electrons with both onsite and neighboring ligand sites \cite{1}. Depending upon the strength of the interaction, many unusual properties such as long-range magnetic order, Kondo effect, heavy-fermion ground state, valence fluctuations, and/or unconventional superconductivity have been observed \cite{jon,greg,k1,k2,111SCa,111SCb,sarrao,Np152SC,k3}. These complex sets of behaviors are well emphasized in uranium compounds with the $UTX$ composition, where $T$ is a $d$-electron transition metal and $X$ stands for a $p$-electron element. The $UTX$  phases crystallize in several different crystal structures, such as the cubic MgAgAs-type, orthorhombic TiNiSi-type, and hexagonal ZrNiAl- and GaGeLi-types \cite{111a,111b}. Depending on the degree of 5$f$-electron localization, many of the  $UPdX$ members show magnetic transitions and features characteristic of strong electronic correlations \cite{rev}. Previous studies showed that uranium and plutonium compounds, UPdSn and PuPdSn, displayed multiple magnetic transitions at low temperatures and characteristics of well localized 5$f$-electrons (which is rare among U and Pu intermetallics) with a small linear specific heat coefficient, $\gamma \sim$5 and 8~mJ/molK$^2$, respectively (see Refs.~\cite{u1,u2,u3,gofPu111}).

While the effect of electronic correlations is relatively well studied in Ce, Yb and U based materials, there is still lack of knowledge on how these collective phenomena impact magnetic, transport, and thermodynamic properties in transuranium intermetallics. Previous studies have shown that NpPdSn behaves as a local-moment antiferromagnet with the N\'eel temperature $T_{N}$=19~K \cite{gofNp111}. In addition, the enhancement of the electronic coefficient of the heat capacity might suggest a strong correlation between 5$f$ and conduction electrons at low temperatures in this material \cite{gofNp111}. Here we present our detailed studies on structural, thermal, transport, and spectroscopic properties of NpPdSn compound. All the results obtained strongly suggest the presence of well localized 5$f$-states in this material. Below $T_{N}$, the electrical and heat properties  are dominated by electron-magnon scattering and an enhanced Sommerfeld coefficient together with characteristic behavior of $MR$(H), $S$(T) and $R_{H}$(T) are all characteristic of systems with strong electronic correlations. We also used a neutron diffraction and $^{237}$Np M\"{o}ssbauer spectroscopy to confirm antiferromagnetic ordering in NpPdSn at 19 K with an ordered moment of about 2.2(2)~$\mu_{B}$/Np along the $c$-axis.

\section {Experimental details}

\begin{figure}[t]
\centering
\includegraphics[width=0.5\textwidth]{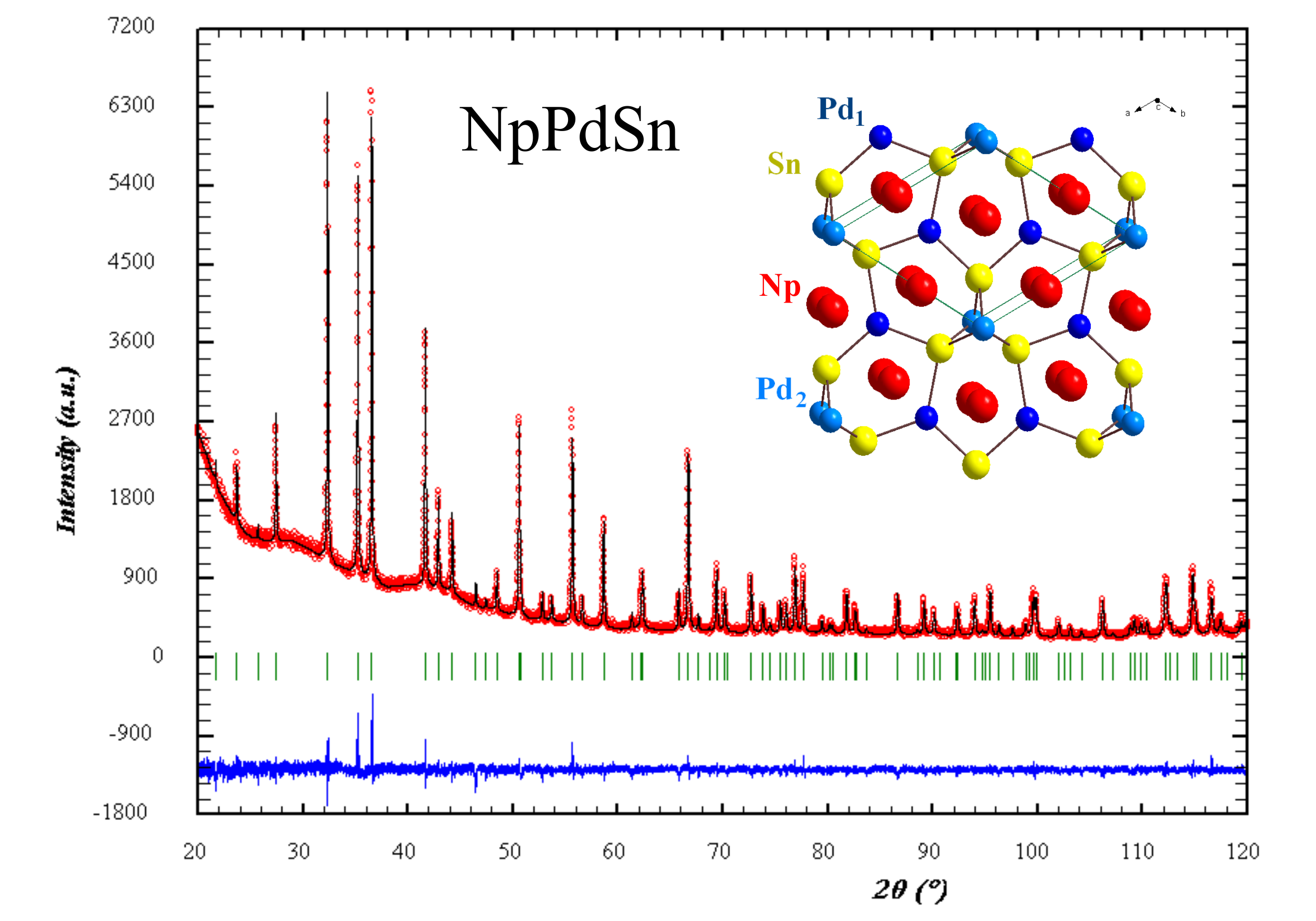}
\caption{(Color) X-ray powder diffraction pattern recorded for NpPdSn at room temperature. All reflections can be attributed to  hexagonal ZrNiAl-type structure shown in the inset. The lower curve represents the difference between the experimental and model results.}\label{x}
\end{figure}

Polycrystalline samples of NpPdSn were synthesized by an arc-melting method. Stoichiometric amounts of high purity Np (99.8\%), Pd (99.999\%), and Sn (99.999\%) were mixed together and arc melted in a Zr-gettered pure argon atmosphere. Small single crystals of NpPdSn were extracted from the ingot to carry out x-ray single crystal diffraction experiments. The powder x-ray diffraction was carried out using a Bruker D8 diffractometer (Cu~K$\alpha_{1}$ radiation). The crystal structure was refined from the single-crystal x-ray data and shown to be hexagonal with the ZrNiAl-type structure (space group $P\bar{6}2m$) with lattice parameters $a$ = 7.5076(6)~\AA~and $c$ = 4.0954(4)~\AA. The atomic coordinates obtained were as follows: Np [3g (0.587(6), 0, 0)], Pd$_{1}$ [1b (0, 0, 0)], Pd$_{2}$ [2c ($\frac{1}{3}$, $\frac{2}{3}$, $\frac{1}{2}$)], and Sn [3f (0.251(2), 0, $\frac{1}{2}$)]. Figure \ref{x} shows the x-ray powder diffraction pattern recorded for polycrystalline NpPdSn. No other peaks than expected for ZrNiAl-type of structure were observed. We also synthesized a non-magnetic analogue of NpPdSn, ThPdSn. Similar to NpPdSn it crystalizes in hexagonal crystal structure but of the GaGeLi-type (space group $P6_{3}mc$) and with lattice parameters $a$ = 4.6602(5)~\AA~and $c$ = 7.6316(4)~\AA. The atomic coordinates obtained are: Th [2a (0, 0, $\frac{1}{4}$)], Pd [2b ($\frac{1}{3}$, $\frac{2}{3}$, 0.427(2))], and Sn [2b ($\frac{2}{3}$, $\frac{1}{3}$, 0.511(3))]. 

Electrical resistance and magnetoresistance were measured from 2-300~K in magnetic fields up to 14~T using the ACT option in a Quantum Design Physical Property Measurement System (PPMS). Four platinum wires were attached on a bar shaped polycrystalline sample using silver paint for a standard resistivity measurement. Heat capacity experiments were carried out using the relaxation method employed in the HC Option in the PPMS. The thermoelectric power was measured from 3 to 300 K using a custom-made setup using pure copper as the reference material. All the preparation and encapsulation were carried in specially designed glove boxes under inert nitrogen atmosphere in order to avoid the radioactive contamination risk of elemental neptunium. The neutron powder diffraction experiment took place at the high resolution diffractometer E9 installed at the BER II reactor of the Helmholtz-Zentrum Berlin. We used 1.3 g of NpPdSn fine powder sample encapsulated in a 4 mm diameter capsule certified for helium leak and enclosed in another double-walled leak-proved aluminum container preventing a contamination of the environment.  The container has been attached to a sample stick and inserted into a $^{4}$He-flow ILL-type cryostat used to reach low temperatures. An x-ray radiogram of the capsule is shown in the inset of Fig.~\ref{n2}. Although a capsule with a small diameter was used, a significant absorption correction to the experimental data had to be applied. In order to ensure secure operation, the interior of the cryostat was permanently monitored using an $alpha$ particle detector attached at the exit port of a close-cycle circuit pumping purified argon gas through the sample space. Any un-tidiness of the capsule would result in a radiation alarm. Two incident wavelengths of $\lambda$ = 2.4 \AA~and 1.79~\AA~selected by graphite and copper monochromators were utilized. A set of $\lambda$/2 filters reducing the contamination of higher-order components in the incident beam to a level below 10$^{-4}$ were used. The E9 diffractometer is equipped with a multi-$^{3}$He-detector covering a large range of diffraction angles enabling thus an effective mapping and detection of all the available diffracted signal. During the experiment we have collected mainly diffraction patterns using the 1.79 \AA wavelength at 25 and 5 K, i.e. above the proposed magnetic phase transition temperature and at the lowest temperature for which the sample capsule is certified (that is well below $T_{N}$). At each temperature we have collected data for ~ 20 hours. The data taken at 25 K served as a basis to determine the crystal structure parameters and the data taken at 5 K were used to identify the signal caused by the anticipated long-range magnetic ordering.  The crystal structure and magnetic structure refinements were carried out with the program Fullprof (part of Winplotr suite \cite{fullprof}). In the refinements, the nuclear scattering lengths $b$(Np) = 10.55 fm, $b$(Pd) = 5.91 fm, and $b$(Sn) = 6.225 fm were used \cite{scatlengths}. In the case of the magnetic structure refinement, we further assumed that Np is in the Np$^{3+}$ state. In this case, the magnetic form factor of can be calculated in a dipole approximation and tabulated in the literature \cite{formfactor1,formfactor2}. The $^{237}$Np M\"{o}ssbauer measurements were performed on a powder absorber with a thickness of 140 mg of Np cm$^{-2}$. The M\"{o}ssbauer source of $^{241}$Am metal (108 mCi) was kept at 4.2 K while the temperature of the absorber was varied from 1.7 K to 25 K. The spectra were recorded with a sinusoidal drive system using conventional methods. The velocity scale was calibrated with reference to the standard absorber NpAl$_{2}$ (B$_{hf}$ = 330 T at 4.2 K).

\section{Results and discussion}

\subsection{Heat Capacity}\label{SH}

\begin{figure}[h]
\centering
\includegraphics[width=0.45\textwidth]{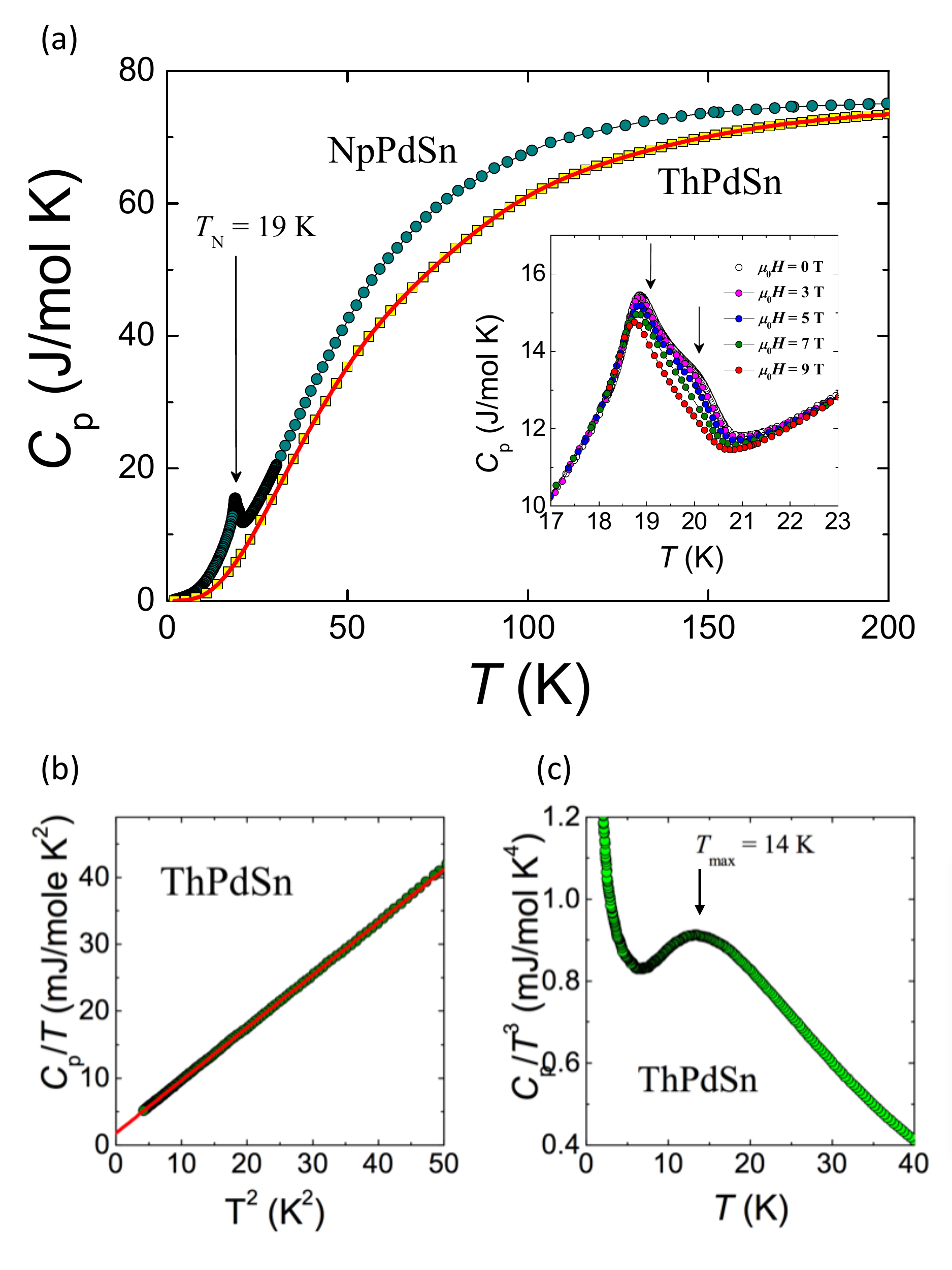}
\caption{(Color online) (a) Temperature dependence of the specific heat of NpPdSn and ThPdSn. For ThPdSn, the solid line represents a fit according to eq.~\ref{fraga} (see the text). The inset show a magnetic field dependence of the specific heat in the vicinity of the N\'eel temperature (lower inset). Arrows mark the phase transitions; (b) the low temperature part of the heat capacity of ThPdSn together with the fit to eq.~\ref{cplow}; (c) the specific heat data of ThPdSn in the form of a $C_{p}/T^{3}$ vs. T plot.}\label{cp}
\end{figure}

Figure~\ref{cp}a shows the temperature dependence of the specific heat, $C_p(T)$, of NpPdSn. At 200 K, the value of $C_p (T)$ is 74.8~J/molK, which is close to the Dulong-Petit value $C_p$ = 3$nR$ = 74.83~J/molK, where $n$ is the number of atoms per molecule ($n$ = 3) and $R$ is the gas constant. A pronounced peak is observed at $T_{N}$ = 19~K due to the antiferromagnetic ordering. The value of this antiferromagnetic transition temperature, $T_{N}$, is consistent with the previous report \cite{gofNp111}. Close inspection of this anomaly (see the inset of Fig.~\ref{cp}a) suggests that there could be a second transition at about 20.2~K, similar to observed in UPdSn and PuPdSn (see Refs.~\cite{u1,u2,u3,gofPu111}). The origin of the second anomaly is unclear and is a subject of on-going investigation. With an application of magnetic field, the peaks shift slightly to lower temperature, consistent with AFM ordering, and faintly lowers its magnitude. 

To estimate the non-lattice contribution to the total specific heat of NpPdSn we assume that the lattice part of the heat capacity of NpPdSn is similar to that in its non-magnetic analogue ThPdSn. Though it is not ideally isostructural, its similar symmetry, atomic mass, and volume per $f.u.$ will approximately match the phonon density of states at lower frequencies in the Debye approximation, which will determine the low temperature behavior \cite{walter}. This material contains no 5$f$-electrons and shows weak diamagnetic properties ($\chi$~=~-2.55 $\times$ 10$^{-5}$ ~emu/mol at room temperature). Fig.~\ref{cp}b shows the temperature dependence of the specific heat of ThPdSn. Below $T$~$\approx$~10~K the $C_{p}(T)$ variation of ThPdSn can be well described (see the inset to Fig.~\ref{cp}b) by the formula:

\begin{equation}
C_{p}(T) = \gamma T + \beta T^{3},\label{cplow}
\end{equation}

with the coefficients $\gamma$~=~2~mJ/molK$^{2}$ and $\beta$ = 7.9~$\times$~10$^{-4}$~J/molK$^{4}$. The electronic contribution to the specific heat is very small, in accordance with the metallic-like behavior of the resistivity of this material (cf. Section~B). From the value of $\beta$ one estimates the Debye temperature $\Theta_{D}~=~(12R\pi^{4}n/5\beta)^{1/3}$ to be about 200~K. In order to describe the temperature variation of the specific heat of ThPdSn in a wider temperature range we use the expression: \cite{fraga}

\begin{equation}
C_{\rm p}(T)~=~\gamma T + \left(1 - k\right)C_{\rm D}(T) + kC_{\rm E}(T)\label{fraga}
\end{equation}

in which the lattice contribution to the specific heat is accounted for by considering both the Debye:

\begin{equation}
C_{\rm D}(T)~=
9nR\left(\frac{T}{\Theta_{\rm D}}\right)^{3}\int^{\frac{\Theta_{\rm D}}{T}}_{0}\frac{x^{4}e^{x}}{\left[e^{x}-1\right]^{2}}dx,\label{deb}
\end{equation}

and the Einstein function:

\begin{equation}
C_{\rm E}(T)~=
3nR\left(\frac{\Theta_{\rm E}}{T}\right)^{2}\left\{e^{\frac{\Theta_{\rm E}}{T}}-1\right\}^{-2}e^{\frac{\Theta_{\rm E}}{T}},\label{ein}
\end{equation}

with the $k$ parameter ensuring the proper quantity of the oscillator modes involved. The least-squares fit of this formula to the experimental data below 100~K (see Fig.~\ref{cp}b) yields the following parameters: $\gamma$~=~3~mJ/molK$^{2}$, $\Theta_{\rm D}$~=~228~K, $\Theta_{\rm E}$~=~75~K and $k$~=~0.15. It is worth to note that the values of the Sommerfeld coefficient and the Debye temperature are very close to those obtained from eq.\ref{cplow}. Also, the Einstein term gives only a small contribution to the total specific heat of ThPdSn (a small value of $k$). Nevertheless, as displayed in the Fig.\ref{cp}c, the Einstein contribution is not negligible as marked by a distinct maximum in the $C_{p}/T^{3}$ vs. $T$ plot at $\Theta_{max}$ = 14~K (marked by the arrow in Fig.\ref{cp}c). This scales well with the Einstein temperature derived from Eq.\ref{fraga} ($\Theta_{E}$$\approx$5$\Theta_{max}$, see Ref.~\onlinecite{sch}).

\begin{figure}[t]
\centering
\includegraphics[width=0.5\textwidth]{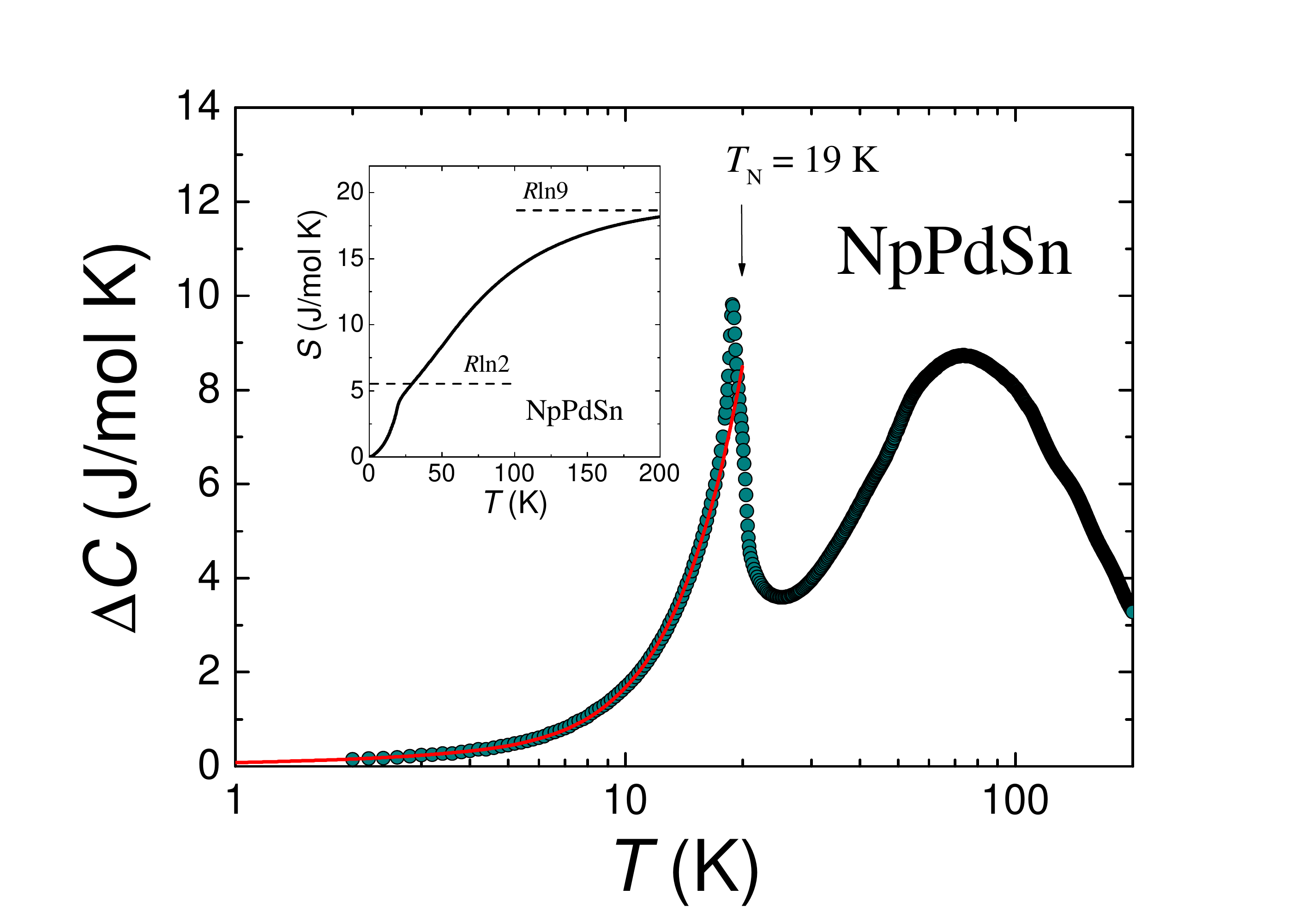}
\caption{(Color online) Temperature variation of the magnetic contribution to the specific heat of NpPdSn in a semilogarithmic scale. The solid red and dashed orange lines are fits as described in the text. The inset shows the temperature dependence of entropy, $S(T)$.}\label{cp2}
\end{figure}

Assuming the phonon part of the specific heat of NpPdSn is similar to ThPdSn, we have derived $\Delta C (T) = C_{p}^{NpPdSn} (T) - C_{ph}^{ThPdSn}(T)$, as shown in Fig.~\ref{cp2}. Low temperature behavior of $\Delta C (T)$ of NpPdSn, below the magnetic phase transition, can be expressed by the equation: 

\begin{equation}\label{delC}
\Delta C (T) = \gamma^* T + C_{mag} (T),
\end{equation}

where the first term represents the electronic contribution due to renormalized quasiparticles and the second term: 
\begin{equation}\label{cmag}
C_{mag} (T) = c \Delta^{7/2} \sqrt{T} e^{-\Delta/T}\bigg[1+\frac{39 T}{20\Delta}+\frac{51}{32}\bigg(\frac{T}{\Delta}\bigg)^{2}\bigg],
\end{equation}

is appropriate for anisotropic antiferromagnetic systems assuming that the dispersion of antiferromagnetic magnons is given by the relation, $\omega = \sqrt{\Delta^{2}+Dk^{2}}$, where $\Delta $ is the gap parameter \cite{cmag1,cmag2}. In the above equation, the coefficient $c$ is given by the relation $c \propto D^{-3/2}$, where $D$ is the spin-wave stiffness. The $\Delta C (T)$ curve below $T_{N}$ can be fitted with eq.~\ref{delC}, shown by a solid red line in Fig.~\ref{cp2}, with the fitting parameters $\gamma^*$ = 82~mJ/molK$^{2}$, $c$ = 2.9 $\times$ 10$^{-5}$ J/molK$^{5}$, and $\Delta$ = 16 K. The low-temperature electronic specific heat of NpPdSn is enhanced as compared to simple metals such as copper or gallium \cite{Cu}. This renormalized Sommerfeld coefficient value can be compared to those reported for UPdIn and UPdSb dense Kondo systems, which are 280 and 62~mJ/molK$^2$ \cite{pm13,pm12}, respectively. The overall shape of the $\Delta C(T)$ curve with a pronounced maximum near 70~K suggests a predominant contribution due to crystalline electric field (CEF). In the hexagonal symmetry and for J = 4 the maximum degeneracy is two, and the nine levels form three singlets and three doublets \cite{CEF}. Detailed inelastic neutron scattering would, however, be necessary to conclude the CEF effect and splitting in NpPdSn. The presence of the doublet CEF ground state in NpPdSn is supported by the magnetic entropy as shown in the inset to Fig.~\ref{cp2}. At high temperatures the entropy is close to the value of $R$ln9, corresponding to the 9-fold degeneracy of the $^5I_{4}$ multiplet. The magnetic entropy around $T_{N}$ reaches close to $R$ln2, corresponding to a doubly degenerated ground state.


\subsection{Electrical Resistivity}

\begin{figure}[t]
\centering
\includegraphics[width=0.5\textwidth]{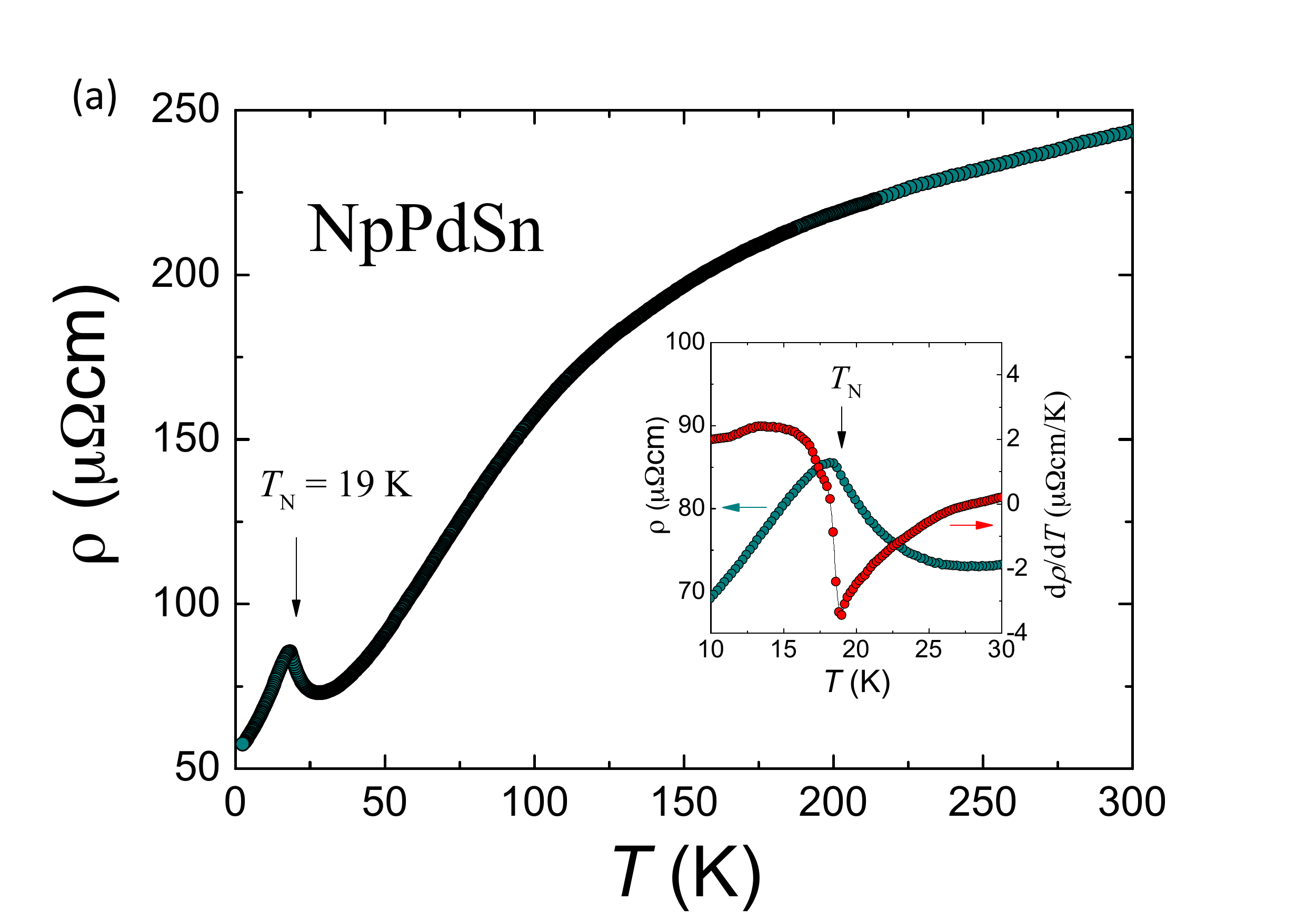}
\includegraphics[width=0.5\textwidth]{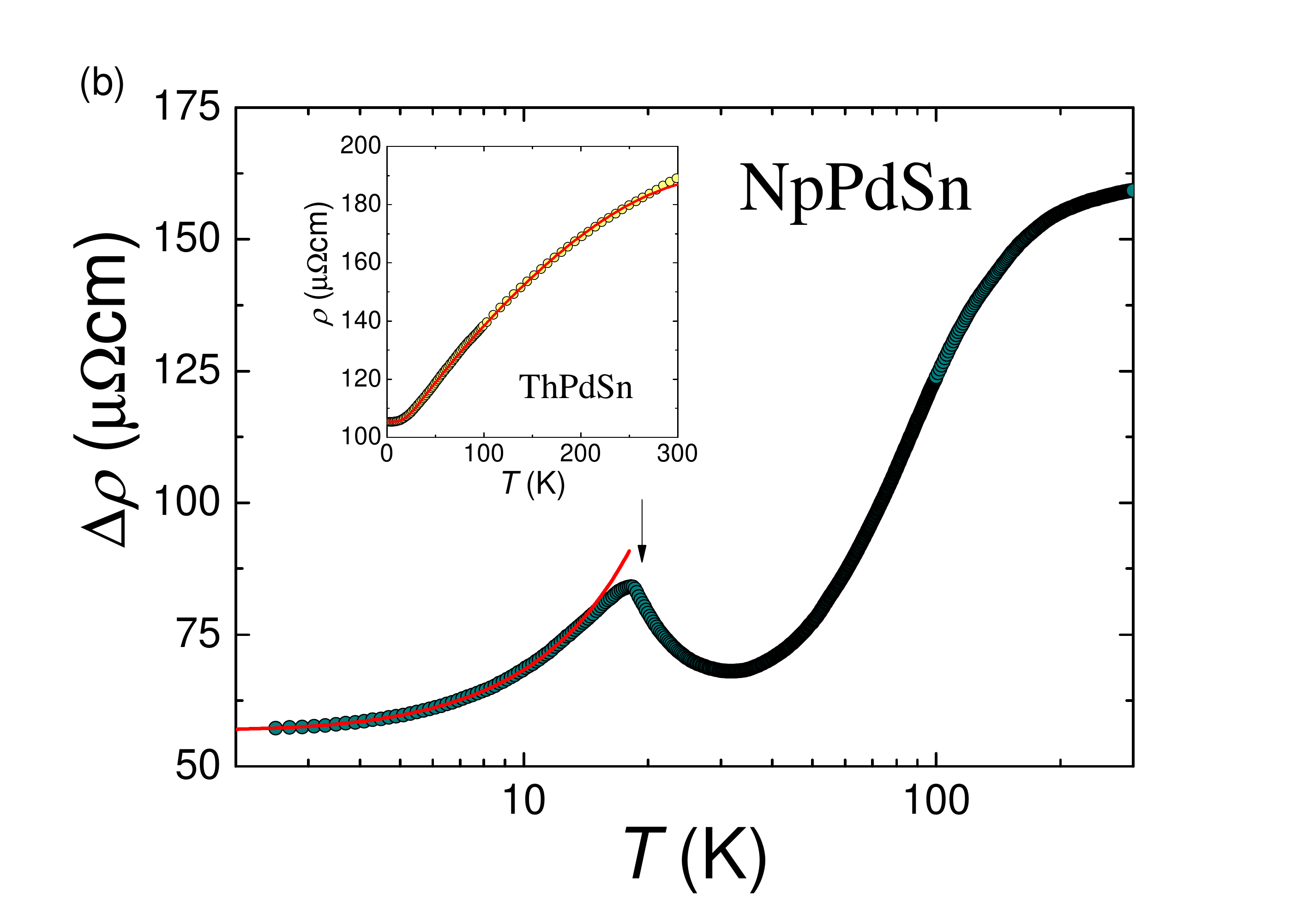}
\caption{(Color online) (a) Temperature dependence of the electrical resistivity of NpPdSn. Inset: low-temperature dependence of the resistivity and the temperature derivative of the resistivity. (b) Temperature variation of the magnetic contribution to the electrical resistivity of NpPdSn. The solid line is fit using the expression given in the text. Inset: the resistivity as a function of temperature of ThPdSn. The solid line is a fit to Bloch-Gr\"{u}neisen-Mott formula (eq.~\ref{r}).}\label{r1}
\end{figure}

Figure \ref{r1}a shows the temperature dependence of the electrical resistivity of NpPdSn. The antiferromagnetic transition appears as a pronounced kink around 19~K. The transition becomes more clear in the temperature derivative of the resistivity as shown in inset to Figure~\ref{r1}a. The temperature variation of resistivity for ThPdSn is shown in the inset to Fig.~\ref{r1}b. The $\rho(T)$ of the Th-based compound is typical for non-magnetic simple metals and may be well described by the Bloch-Gr\"{u}neisen-Mott formula:

\begin{equation}
\rho(T) =
\rho_{0}+4R\Theta_{R}\left(\frac{T}{\Theta_{R}}\right)^{5}\int^{\frac{\Theta_{R}}{T}}_{0}\frac{x^{5}dx}{\left(e^{x}-1\right)\left(1-e^{-x}\right)}-KT^{3},\label{r}
\end{equation}

\begin{figure}[t]
\centering
\includegraphics[width=0.5\textwidth]{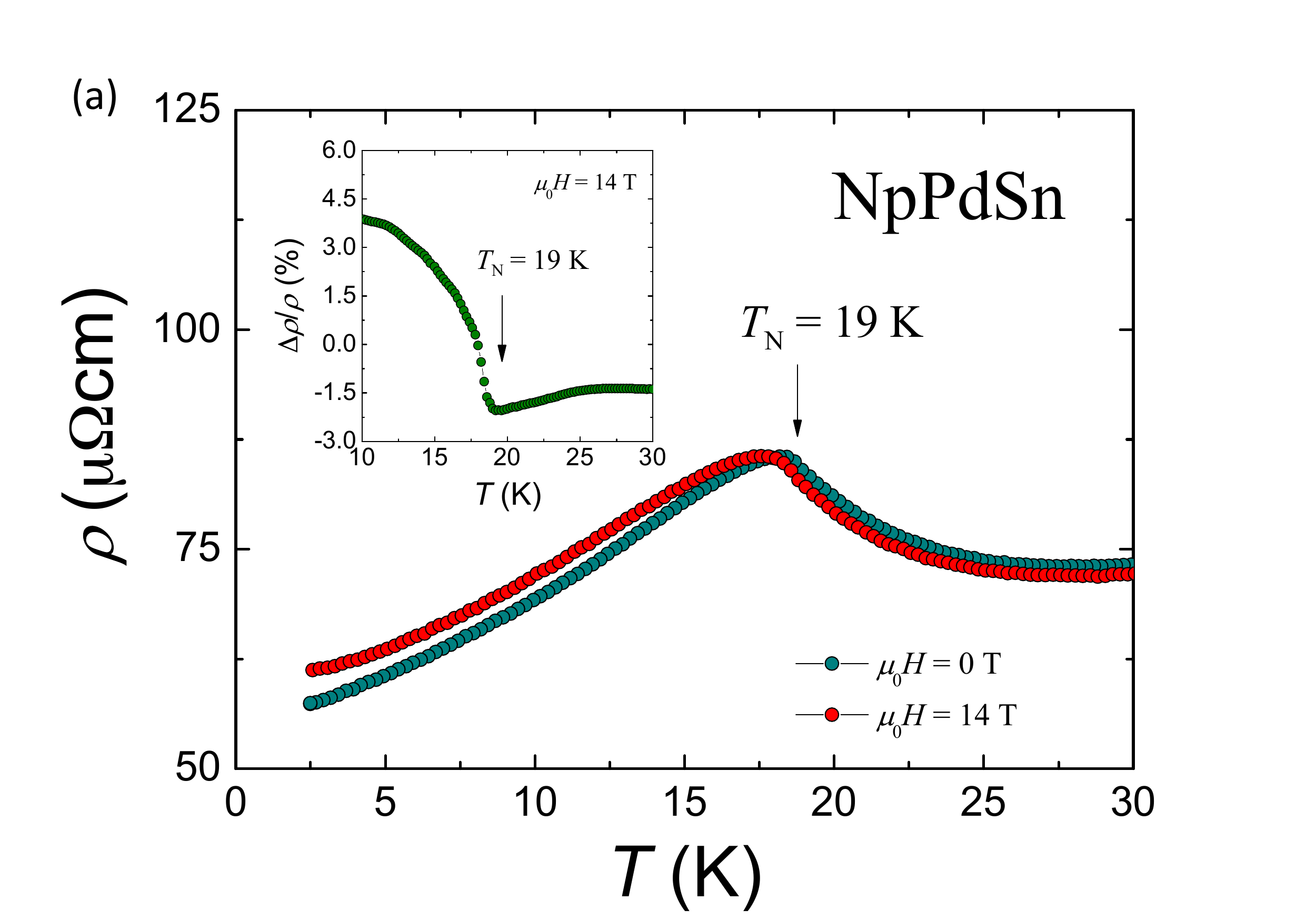}
\includegraphics[width=0.5\textwidth]{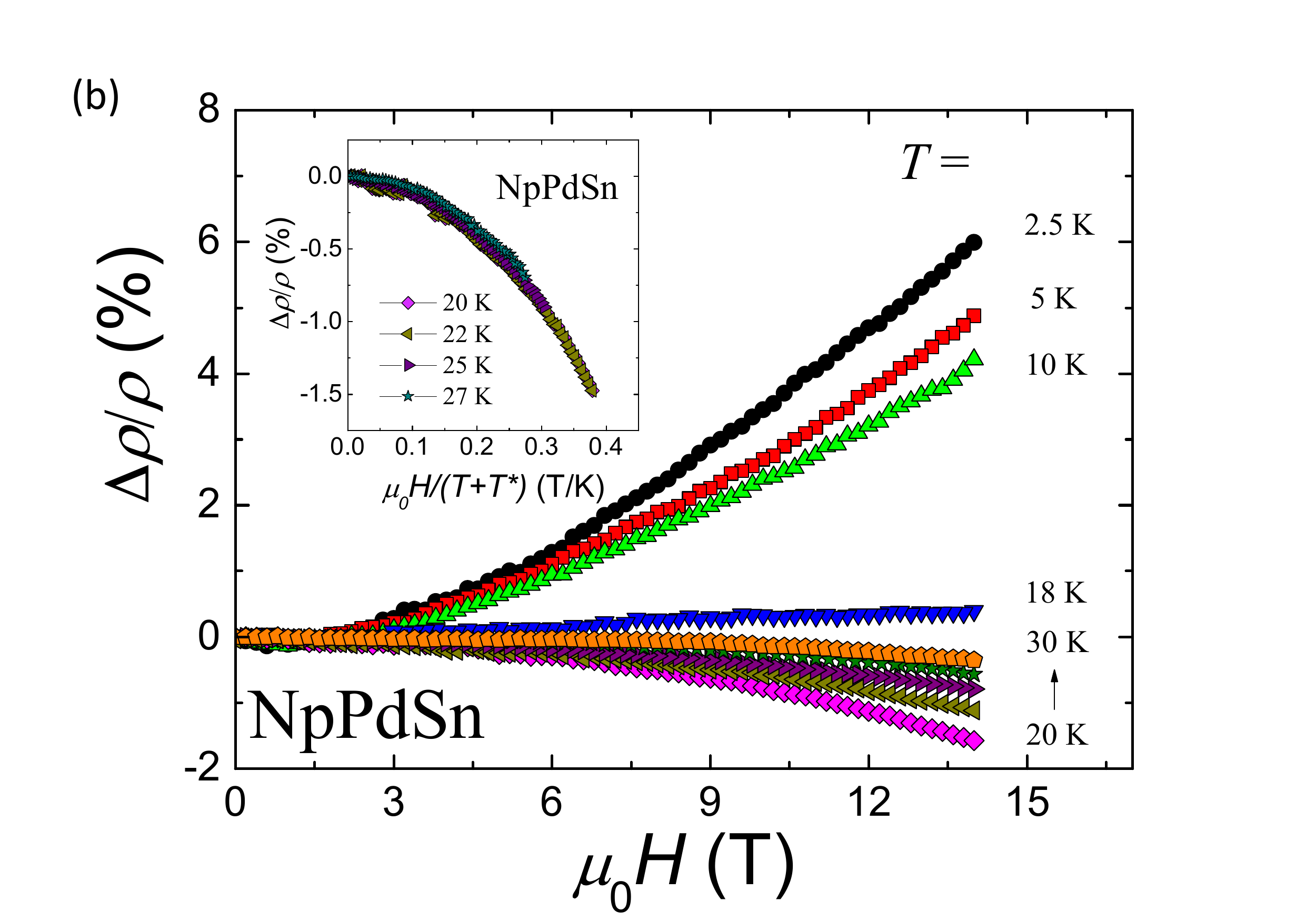}
\caption{(Color online) (a) The temperature dependence of low temperature electrical resistivity measured in 0 and 14~T. The inset shows the temperature dependence of magnetoresistivity taken at 14~T (b) Field dependence of the magnetoresistivity of NpPdSn taken at different temperatures. Inset: the magnetoresistivity of NpPdSn as a function of $\mu_{0}H/(T+T^{*})$ (see the text).}\label{r2}
\end{figure}

where $\rho_{0}$ is the residual resistivity and the second and third terms describe electron-phonon scattering and Mott's $s-d$ interband electron scattering. A least squares fit of this expression to the experimental data of ThPdSn yields the parameters: $\rho_{0}$~=~105~$\mu\Omega$cm, $R$~=~0.4~$\mu\Omega$cm/K, $K$~=~1.1~$\times$~10$^{-6}$~$\mu\Omega$cm/K$^{3}$,  and $\Theta_{R}$~=~125~K. The latter parameter, $\Theta_{R}$, may be considered as an approximation of the Debye temperature $\Theta_{D}$ since $\Theta_{R}$ contains also some contribution related to electron-electron interactions \cite{el}.

While ThPdSn behaves as a non-magnetic simple metal, the temperature dependence of the resistivity of NpPdSn with characteristic increase at low temperatures, as well as its magnitude, is similar to ones observed in Ce- or U-based systems with strong electronic correlations (cf. Refs.~\cite{k1,k2,k3}). Assuming the phonon contribution of the resistivity of NpPdSn is the same as in its non-magnetic counterpart ThPdSn, we have estimated the magnetic contribution to electrical resistivity by using the formula $\Delta \rho(T) = \rho(T)^{NpPdSn} - \rho(T)^{ThPdSn}_{ph}$ as shown in a semilogarithmic scale in Figure~\ref{r1}b. The shape of the $\Delta \rho(T)$ curve is characteristic of Ce-, U-, or Pu-based antiferromagnetic systems with crystal-field interactions (cf. Refs.~\cite{k1,pi,ka,cc}). 

The temperature variation of resistivity below the N\'eel temperature in $\Delta \rho(T)$ can be described by a model that takes into account the scattering of conduction electrons by the antiferromagnetic spin waves with a gap in the magnon spectrum. Similar to the heat capacity analysis, the electrical resistivity of NpPdSb may be expressed by \cite{rmag1,rmag2}:

\begin{equation}\label{rhot}
\rho (T) = \rho_{0} + A \Delta^{2}\sqrt{\frac{k_{B}T}{\Delta}}e^{-\Delta/k_{B}T}\bigg[1+\frac{3\Delta}{2k_{B}T}+\frac{2}{15}\bigg(\frac{\Delta}{k_{B}T}\bigg)^{2}\bigg],
\end{equation}

where $A \propto D^{-3/2}$. The solid red line in figure \ref{r1}b shows the fitting of eq.~\ref{rhot} to the $\Delta \rho (T)$ data below the N\'eel temperature with the least square fitting parameters $\rho_{0}$~=~56.8~$\mu\Omega cm$, $A$ = 0.17~$\mu\Omega cm/K^{2}$ and $\Delta = 7$~K.

Figure~\ref{r2}a shows low-temperature $\rho(T)$ curves measured at magnetic fields of 0 and 14 T. As seen, magnetic field effect on electrical resistivity is not large, being of the order of 6~\% at 2.5~K. The temperature dependence of the transverse magnetoresistivity (i.e. $\mu_{0}H \perp i$), defined as $\frac{\Delta \rho}{\rho} = \frac{\rho (\mu_{0}H) - \rho (0)}{\rho (0)}$ and measured in a magnetic field of 14~T is shown in the inset of Fig.~\ref{r2}a. The $\frac{\Delta \rho}{\rho}$(T) is positive in the ordered state, shows a distinct minimum at the N\'eel temperature and then becomes negative. Such a behavior is typical for antiferromagnetic materials and was theoretically studied by Yamada and Takeda (cf. Ref.~\cite{yt}). Figure~\ref{r2}b displays magnetoresistivity isotherms measured vs. magnetic field at temperatures below and above $T_{N}$. In the ordered state the $\frac{\Delta \rho}{\rho}(\mu_{0}H)$ curves follow the behavior expected for antiferromagnets. Interestingly, above N\'eel temperature, the magnetoresistivity curves have a shape characteristic of systems where week incoherent Kondo-type interactions are suppressed by applied magnetic fields. It has been shown that in these systems the  $\frac{\Delta \rho}{\rho}(\mu_{0}H)$ curves may be scaled (within the Bethe ansatz approximations of the Coqblin- Schrieffer model \cite{p1}) using the relation \cite{p2}:

\begin{equation}
\frac{\Delta\rho}{\rho}\left(\mu_{0}H\right)=f\left(\frac{\mu_{0}H}{T+T^{*}}\right)\label{mr}.
\end{equation}

This formula reflects the fact that for the single impurity the physics of the Kondo impurity is dominated by only one energy scale. In the case of NpPdSn the best overlap of the magnetoresistivity isotherms was obtained for a characteristic temperature $T^{*}$ of 20.5~K which might be considered as a measure of the Kondo energy scale (see Fig.~\ref{r2}b).

\subsection{Seebeck and Hall effects}

Figure~\ref{tfig}a shows the temperature dependence of the Seebeck coefficient $S(T)$ of NpPdSn. At room temperature the Seebeck coefficient is -10 $\mu V/K$ and the magnetic phase transition appears only as a small kink at 19 K in $S(T)$. The transition is more visible and appears as a dip in the temperature derivative of $S(T)$ as shown in figure~\ref{tfig}a inset. The temperature variation of the Seebeck coefficient of NpPdSn is similar to $S(T)$ observed is correlated systems \cite{t1,t2,t3,t4,t5,t6,t7}. To explain this behavior, a simple phenomenological model is proposed which takes into account scattering processes of conduction electrons by a narrow $5f$ quasiparticle band. Considering the Lorentzian form of the band the temperature variation of $S(T)$ can be expressed as \cite{t1,t2}:

\begin{figure}[t]
\centering
\includegraphics[width=0.5\textwidth]{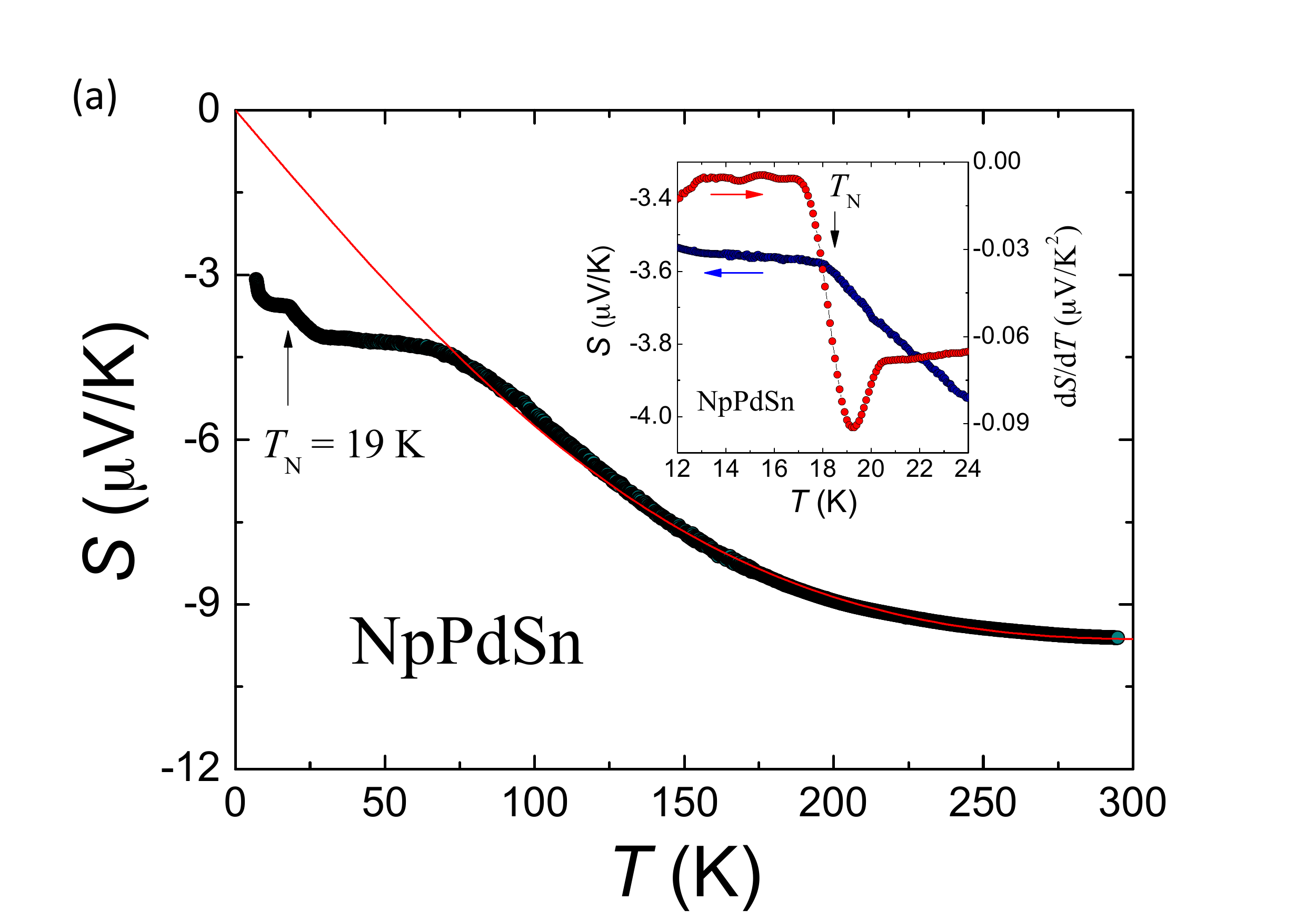}
\includegraphics[width=0.5\textwidth]{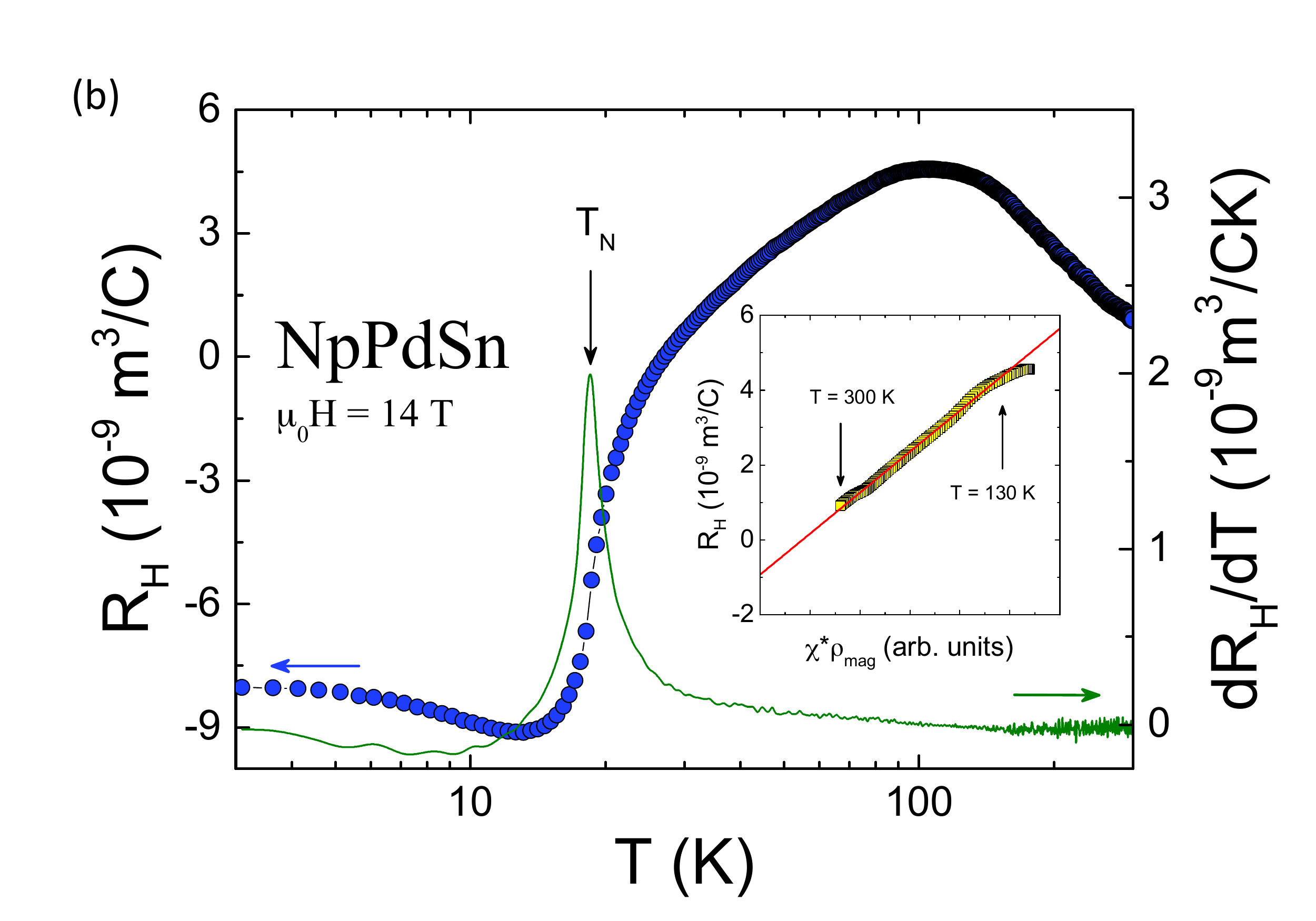}
\caption{(Color online) (a) Temperature dependence of the thermoelectric power of NpPdSn. The solid line is a fit of eq.~\ref{t} to the experimental data. Inset: the Seebeck coefficient and temperature dependence of the derivative of the thermopower in the vicinity of the magnetic phase transition. (b) Temperature dependence of the Hall coefficient and derivative of $R_{H}$, d$R_{H}$/dT of NpPdSn measured in magnetic field $\mu_{0}H$~=~14~T. Inset: $R_{H}$ vs. $\chi^{*}(T)\rho_{mag}$ where $\chi^{*} = \frac{\chi(T)}{C}$ (see the text). The solid line is a fit of eq.~\ref{fert} to the experimental data (see the text).}\label{tfig}
\end{figure}

\begin{equation}
S(T)=\frac{2\pi^2 k_B}{3|e|}\frac{\epsilon_{F} T}{\frac{\pi^{2}T^{2}}{3}+\epsilon^{2}_F + \Gamma^2_F}\label{t}
\end{equation}

where $\Gamma$ denotes its width and $\epsilon_{F}$ is its position with respect to the Fermi energy. Equation~\ref{t} is fitted to the temperature dependence of the Seebeck coefficient above 75 K, as shown by a solid red line in Fig.~\ref{tfig}a. The least squares fitting produces the optimized parameters $\epsilon_F$~=~-3 meV and $\Gamma_F$~=~47~meV. The values of those parameters (and their ratio $\Gamma_F$/$\epsilon_F$$\sim$15.6) are comparable with those obtained for other strongly correlated $5f$ electron compounds (cf. Refs.~\cite{t3,t4,t5,t6,t7,kgnew}). It should be noted that the model described above neglects several interactions that may be present in NpPdSn below $\sim$70 K. More advanced approaches (e.g. taking into account the effect of crystal electric field) should be considered (see Refs.~\cite{z1,z2,z3,z4,z5,z6}).

Figure~\ref{tfig}b shows the temperature dependence of the Hall coefficient $R_{H}(T)$ of NpPdSn taken at 14~T. At room temperature the Hall coefficient is positive and of the order of 0.8$\times$10$^{-9}$m$^{3}$C$^{-1}$. With decreasing temperature, it increases and forms a broad maximum at around 100~K. With further lowering of the temperature the Hall coefficient rapidly decreases, changes sign at around 25~K and form a minimum at 13~K. The overall behavior of $R_{H}(T)$ is similar to the ones obtained for cerium-based correlated \cite{pi,fert,17,39}. According to the theory by Fert and Levy, the temperature dependence of the Hall coefficient may be expressed as \cite{fert}:

\begin{equation}
R_{H}(T)= R_{0}+\gamma_{l}\frac{\chi(T)}{C}\rho_{mag}(T),\label{fert}
\end{equation}

where the first term describes the ordinary Hall effect due to Lorenz motion of carriers and residual skew scattering by defects, while the second term is related to intrinsic skew scattering. In this formula $\chi(T)$ is the magnetic susceptibility, whereas $\rho_{mag}(T)$ is the magnetic contribution to the electrical resistivity. $C$ is the Curie-Weiss constant and parameter $\gamma_{l}$ is a T-independent coefficient related to the phase shift (see Ref.~\cite{fert}). In the case of NpPdSn, least squares fitting of eq.~\ref{fert} to the experimental data in the temperature range 110--300~K (see the inset to Fig.~\ref{tfig}b) resulted in the values $R_{0}$~= -4.2$\times$10$^{-9}$m$^{3}/$C and $\gamma_{l}$~=~1.1~K/T. A straightforward one band model provides an estimate for the concentration of free electrons to be 1.5~$\times$~10$^{21}$cm$^{-3}$, which should be considered as the upper limit of the actual carrier concentration in NpPdSn. The negative sign of $R_{0}$ agrees with a negative sign of Seebeck coefficient and may indicate that electrons are the main carriers in heat and electrical conduction in NpPdSn.

\subsection {Neutron diffraction}\label{ND}

\begin{figure}[t]
\centering
\includegraphics[width=0.5\textwidth]{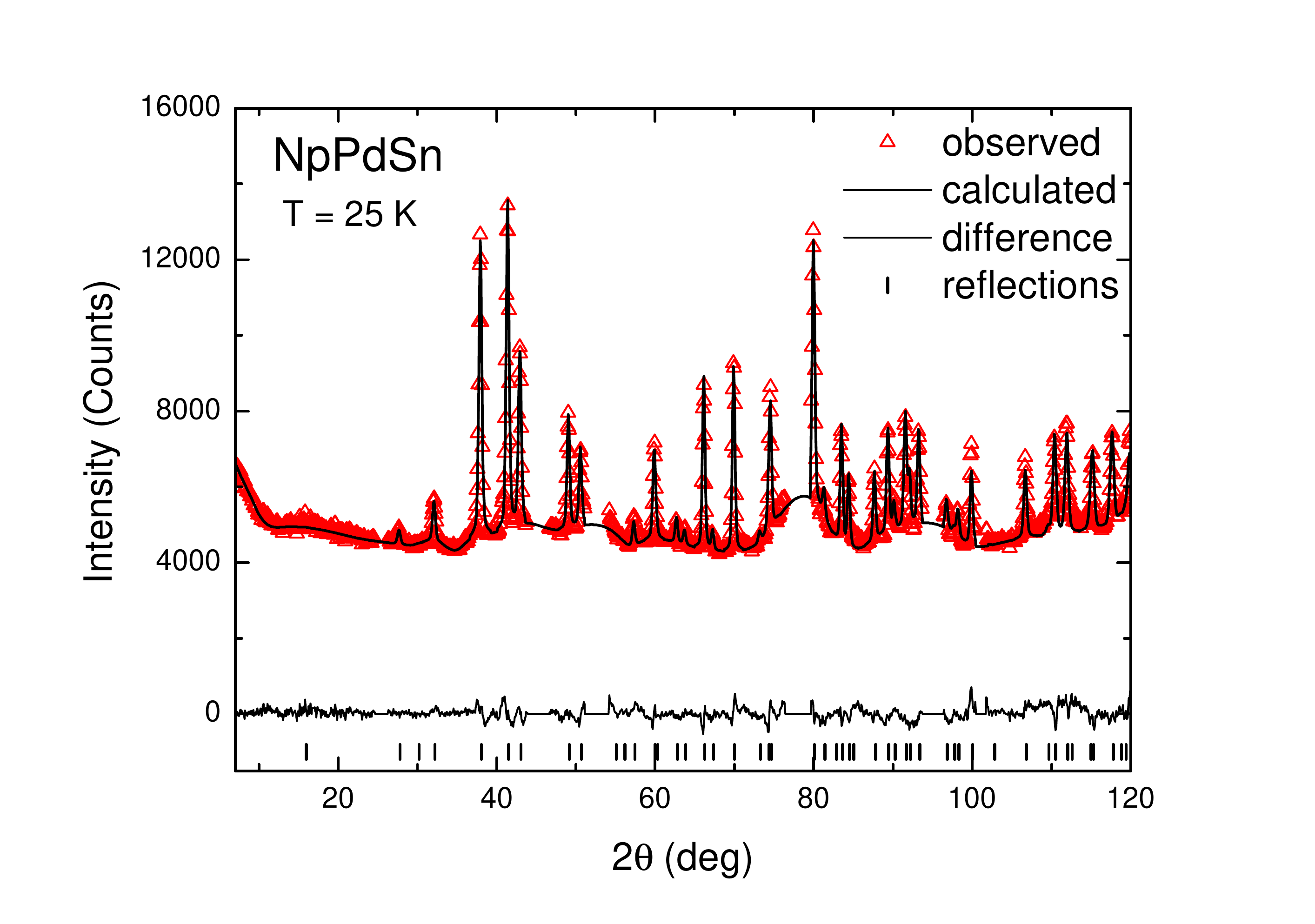}
\caption{(Color online) Portion of diffraction patterns of NpPdSn powder sample taken on E9 at 25 K (red open points). The full line through observed points id the best fit. The line at the bottom is the difference and positions of Bragg reflections are marked by ticks at the bottom. Parts of the spectra (without experimental data) were excluded from the refinement due to heavy sample environment contributions.}\label{n1}
\end{figure}

From the pattern obtained at 25~K, we were able to determine the basic structural parameters that are listed below that are in good agreement with the structure reported from x-ray experiments. In the course of the refinement we have checked for a possible random occupation or site mixing between Np (site 3g) and Pd (sites 1b and 2c, respectively) and Sn (site 3f) atoms. No such effects were found. Nevertheless a slight deficiency of the Pd atoms outside the Np-Pd1 layer and of the Sn atoms has been found. Also, a small preferential orientation was identified. The best fit ($\chi^{2}$ = 4.30, $R_{B}$ = 7.80) to data in the range of diffraction angles up to 120 deg. is shown in Fig. \ref{n1} and the numerical values are listed in Table \ref{t1}. 

\begin{table}[b]
\caption{ Structural parameters of NpPdSn at 25 K}
\centering
\resizebox{0.4\textwidth}{!}{\begin{tabular}{l l l l l l l}
\hline\hline
Atom & Site & x & y & z & B(\AA) & occupancy \\  [0.5ex]
\hline
Np&3g&0.5876(5)&0&0.5&0.3(1)&1.00 (fixed)\\
Pd1&1b&0&0&0.5&0.5(1)&1.01(3)\\
Pd2&2c&1/3&2/3&0&0.4(2)&0.95(4)\\
Sn&3f&0.2471(7)&0&0&0.6(1)&0.91(3)\\ [1ex]
\hline
\end{tabular}}

Unit cell parameters: $a$ = 7.4866(4) \AA, $c$ = 4.0759(2) \AA
\label{t1}
\end{table}

\begin{figure}[t]
\centering
\includegraphics[width=0.5\textwidth]{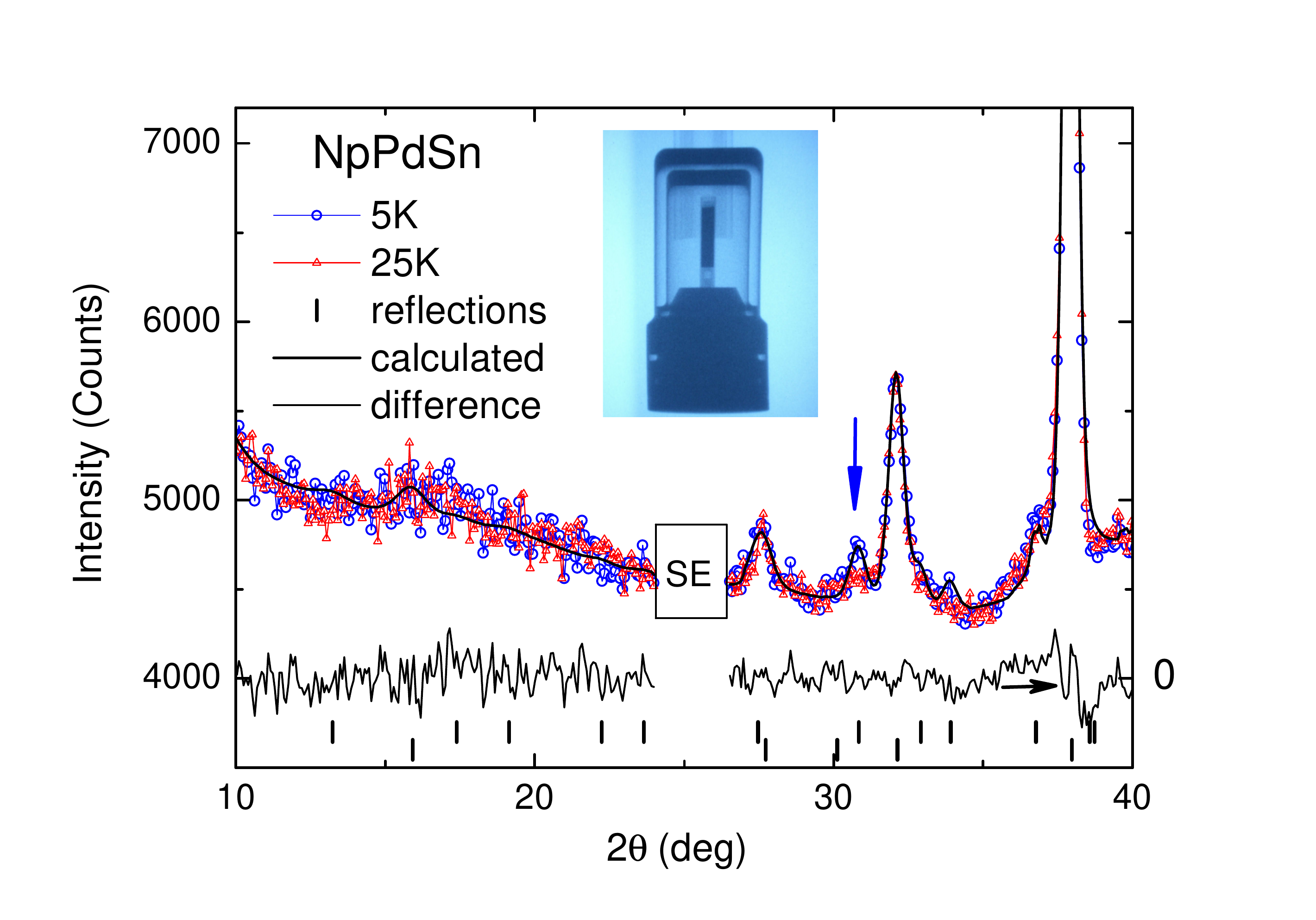}
\caption{(Color online) Portion of diffraction patterns taken on E9 at 5 K (blue curve) and at 25 K (red curve). The strongest Bragg reflection that is present only in the 5 K pattern and absent at 25 K is marked by blue arrow. SE denotes diffraction angles excluded from refinement due to parasitic sample environment signal. Inset: x-ray picture of the NpPdSn sample (total mass 1.3 g) encapsulated in a 4 mm diameter capsule and enclosed in double-walled container.}\label{n2}
\end{figure}

The subtraction method applied to the diffraction pattern collected well below the magnetic phase leads to detection of Bragg reflections that appear due to magnetic order. Rather small magnetic moments could be normally detected. In the case of NpPdSn it appears that the 5 and 25~K patterns are nearly identical. The resulting differential pattern reveals a presence of several very weak additional Bragg reflections that could be associated with the magnetic order (see Fig.~\ref{n2}). Most of them are at the level of statistical uncertainty. However, the strongest additional Bragg reflection that has the $d$- spacing of 3.37~\AA~and appears around 2$\theta$ = 30.8 degrees (in Fig.~\ref{n2} denoted by a thick blue arrow) does not coincide with any nuclear Bragg reflection. Therefore we conclude that the magnetic structure of NpPdSn is indeed of an antiferromagnetic type. Indexing of the additional reflections is ambiguous as several commensurate \textbf{q}$_{1}$ = (0.25, 0.25, 0) and \textbf{q}$_{2}$ = (0.25, 0, 0.5) and incommensurate (e.g. \textbf{q}$_{0}$ = (0.5, 0, 0.428)) propagation vectors can account for observed reflections. Especially the commensurate propagation vectors seems to be more realistic as generally magnetic structures tend to be commensurate at low temperatures, consisting of equal-size moments. The latter feature is in view of M\"{o}ssbauer spectroscopy results (see below), however, not expected. Instead, we look for a magnetic structure that allows for a coexistence of fully developed and negligible Np moments. For all the propagation vectors that lead magnetic Bragg reflection positions compatible with observed pattern at 5 K, we have generated all possible moment configurations using a symmetry group analysis as developed by Bertaut \cite{bertaut}. It appears that nearly all the models allow for a coexistence of big and negligible Np moments. For instance, for \textbf{q}$_{0}$ = (0.5, 0, 0.428) there are two symmetrically coupled moments with the third detached, leading to a possibility that the latter one remains zero. However, the remaining moments would need to be sine-wave modulated along the $c$ axis leading to an infinite number of moment magnitudes. Such a magnetic structure has been observed e.g. in CePdAl \cite{donni}, where one of the Ce moments does not order down to dilution temperatures. Group analysis for \textbf{q}$_{1}$ = (0.25, 0.25, 0) lead to similar splitting of the three sites. Here, however, the moments are modulated commensurately and can reach only a finite number of different magnitudes. For \textbf{q}$_{2}$ = (0.25, 0, 0.5) the three moments split in three independent positions. Also here the moments can reach only a limited number of values. Let us assume that the magnetic structure is described by  such a propagation vector. It appears there are only two one-dimensional irreducible representations, with all Np moments either oriented along the hexagonal axis or perpendicular to it. 

\begin{figure}[t]
\centering
\includegraphics[width=0.5\textwidth]{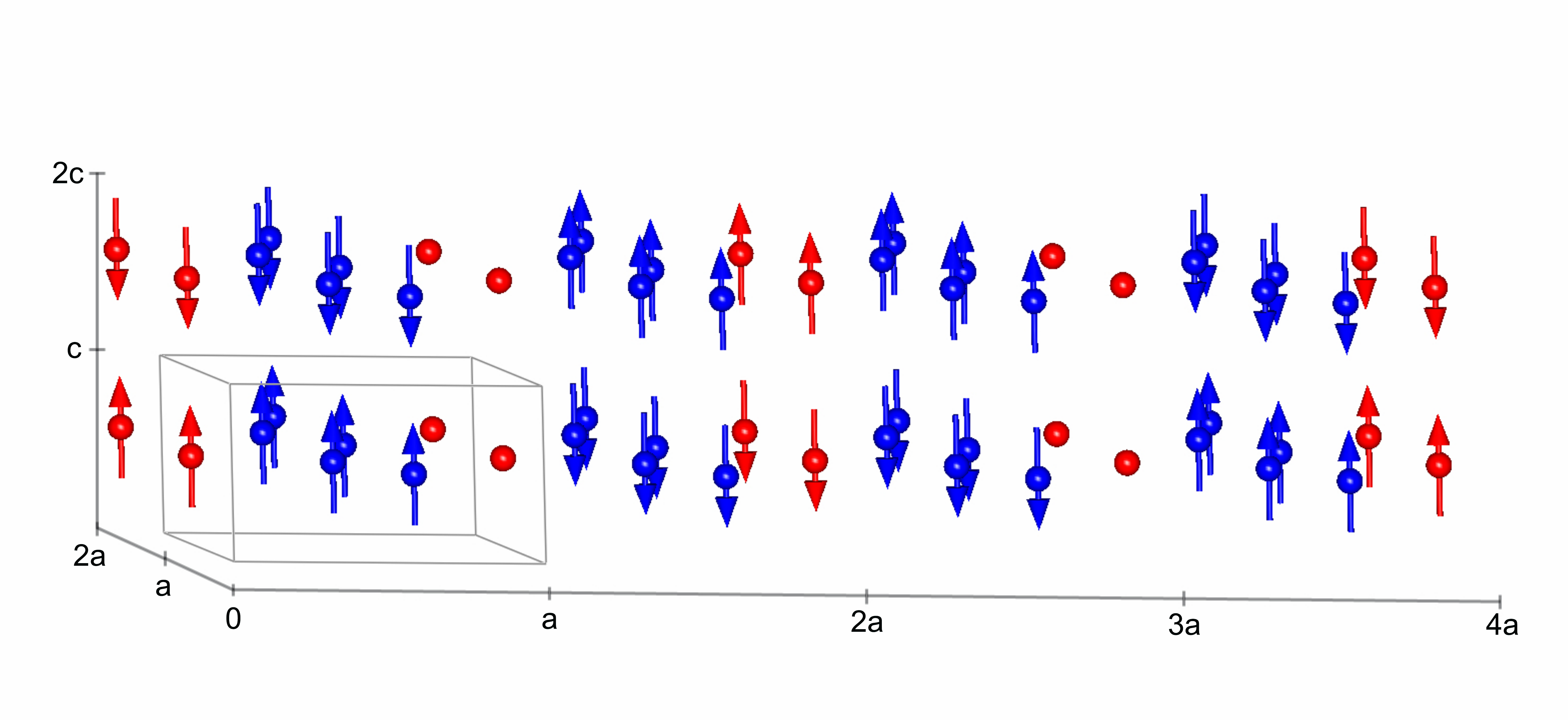}
\caption{(Color online) A schematic representation of a possible antiferromagnetic structure of NpPdSn (only magnetic moments are shown). Np magnetic moments are oriented along the c axis with stacking + -- along the c axis. The stacking along the a axis is for two of the three Np moments (blue) of the + + -- -- type. The remaining moment (red) has along this direction the + 0 -- 0 stacking.}\label{n3}
\end{figure}

We have performed a number of fits in which we kept the moment at different sites either fixed to the identical value or let them  be independent, pointing along and perpendicular to the hexagonal axis. The best agreement has been obtained for model with moments pointing along the $c$ axis, in which two of the Np moments are equivalent and the third one (at 0 0.58 0.5) has a phase shift with respect to the first two of 0.125$\pi$. As a consequence, moving along the a axis, every second moment at this position is equal to zero (see Fig. \ref{n3}). The remaining two moments show a typical + + -- -- coupling along the a axis forming ferromagnetic chains along the other equivalent a axis direction. It appears that the fully developed moment (at site 0, 0.58, 0.5) is situated between the two fully developed moments at the remaining sites and the vanishing moment between positively and negatively oriented moments.  As can be seen from the  Fig. \ref{n2}, the agreement between the observed data and calculated profile for this magnetic structure is rather good with $\chi^{2}$ = 2.30 and $R_{M}$ = 15.92. Refined Np magnetic moment amounts to about 2.2(2)~$\mu_{B}$/Np. The value of ordered moment is slightly reduced if compared to the theoretical value 2.4 $\mu_{B}$ for Np$^{3+}$ ion. This may imply an influence of the CEF effect and/or Kondo-type screening of the conduction electrons on the localized magnetic Np moments in NpPdSn. It has to be noted that this solution is only one of possible configurations. The data are rather limited and the details of the magnetic structure remain unclear. However, the magnetic structure described here has one big advantage: it consists of commensurately modulated Np moments and allows for a coexistence of fully developed and vanishing moments. The ratio between the two groups is 5:1, a value that is close to ratio found in M\"{o}ssbauer spectroscopy (see section \ref{MS} below). At the same time, it allows for temperature changes as one can expect that the moment magnitudes at various sites gradually develop with temperature in a different manner changing thus the ratio between fully developed and vanishing moments. NpPdSn is not the first Np-based compound studied. Some time ago, Javorsk\'{y} et al. reported on the neutron diffraction of NpIrSn \cite{jav}. Although at two different temperatures, two clearly different sets of magnetic reflections have been identified, no unambiguous indexation could be found.

\subsection {$^{237}$Np M\"{o}ssbauer spectroscopy}\label{MS}

\begin{figure}[t]
\centering
\includegraphics[width=0.5\textwidth]{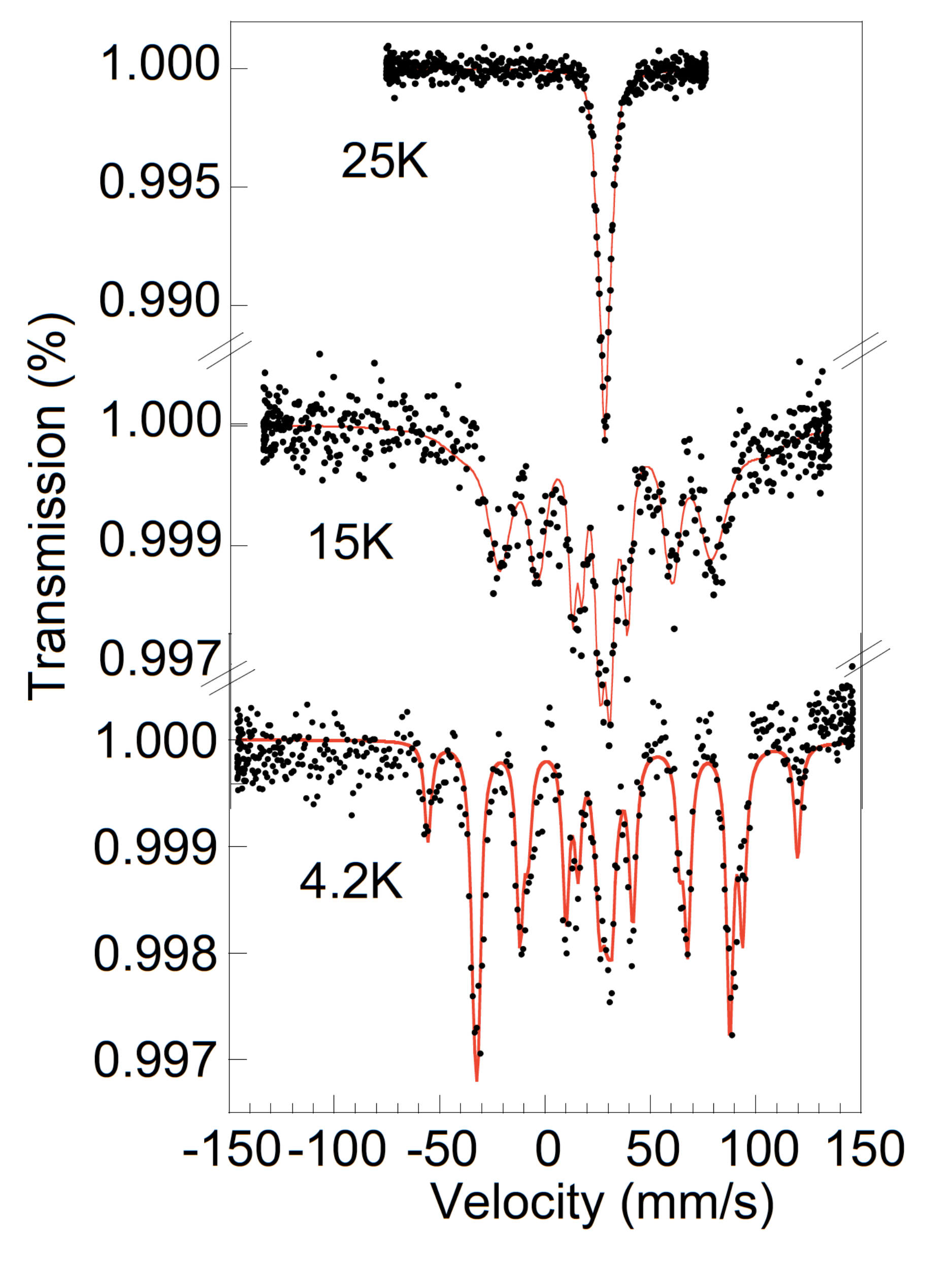}
\caption{(Color online) $^{237}$Np M\"{o}ssbauer spectra recorded at typical temperatures in the paramagnetic (T = 25 K) and antiferromagnetic (T = 4.2 and 15 K) states. The solid (red) lines represent the fits obtained by solving the complete Hamiltonian for the hyperfine interactions.}\label{M1}
\end{figure}

\begin{figure}[t]
\centering
\includegraphics[width=0.5\textwidth]{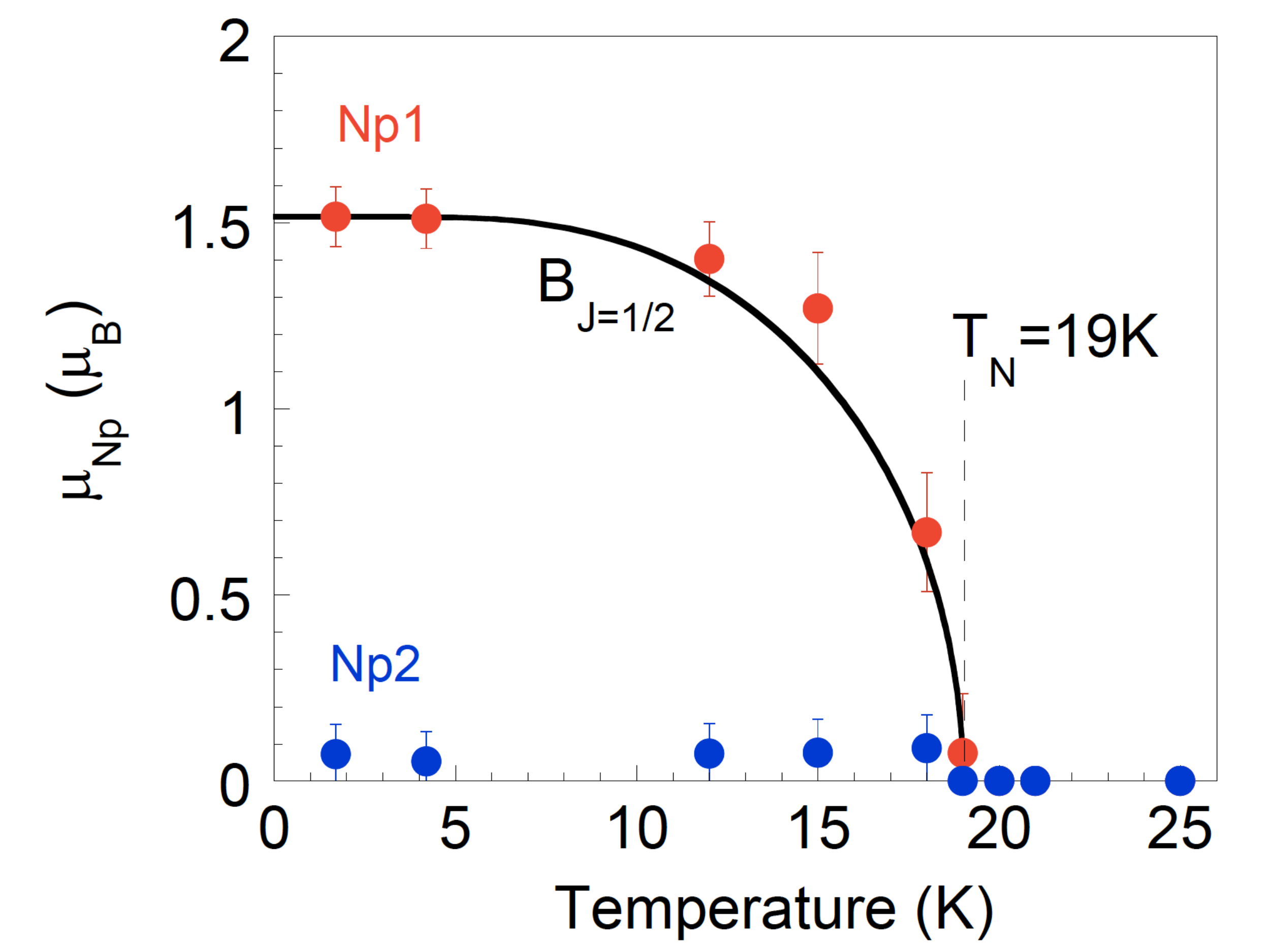}
\caption{(Color online) Temperature dependence of the ordered magnetic moment carried by Np1 (large moment) and Np2 (small moment) atoms in NpPdSn. The solid line represents the Brillouin function $B_{J}$ for J = 1/2.}\label{M2}
\end{figure}

Figure \ref{M1} shows  M\"{o}ssbauer spectra recorded at different, typical temperatures. In the paramagnetic state (T = 25 K), we observe a single site of neptunium, in agreement with the crystal structure, with no ordered magnetic moment, a small quadrupolar interaction ($\mid e^{2}qQ\mid$ = 14.1 mm/s) and an isomer shift ($\delta$IS = 14.5 mm/s versus NpAl$_{2}$) indicative of a Np$^{3+}$ charge state (electronic configuration 5f$^{4}$). In the antiferromagnetic state (T = 4.2 K and 15 K), the M\"{o}ssbauer spectrum is split by the hyperfine magnetic field that raises the degeneracy of the nuclear levels, confirming the ordering of magnetic moments carried by neptunium atoms. Contrarily to the paramagnetic state, the pattern cannot be accounted for by a single neptunium site: below the N\'eel temperature, 2 sites with different hyperfine fields (H$_{hf}$ = 326 T and 16 T at T = 1.7 K, not shown) but identical isomer shift and quadrupolar interaction parameters are observed. This means that all neptunium atoms have the same crystallographic environment but do not carry the same magnetic moment: at low temperature, large moments ($\mu_{Np1}$ = 1.52(8) $\mu_{B}$) coexist with vanishing moments ($\mu_{Np2}$ = 0.07(6) $\mu_{B}$) in the ratio 6:1. This ratio, obtained at T = 4.2 K, compares well with the one inferred at T = 5 K from neutron diffraction (5:1), in particular when taking into account that this ratio decreases with increasing temperature, as illustrated at T = 15 K (3:1) where the central part of the spectrum (small moment) clearly increases at the expense of the external lines (large moment). The large moment $\mu_{Np1}$ is slightly smaller than the one found by neutron diffraction but still in reasonable agreement, this suggests that the correct value is probably between the upper part of the error bar for the Mössbauer value and the lower part of the error bar for the neutron value. It is interesting to mention that the quadrupolar interaction parameter observed in the antiferromagnetic state ($e^{2}qQ$ = -14.1 mm/s) has the same absolute value as in the paramagnetic state, which indicates that the magnetic moments are collinear and oriented along the direction of the main component of the principal axis of the electric field gradient $V_{zz}$, i.e. along the c-axis of the hexagonal crystal, in agreement with the neutron experiments (see section \ref{ND}). The temperature dependence of the magnetic moments Np1 and Np2 is represented on Figure \ref{M2} and agrees well with a N\'eel temperature T$_{N}$ = 19 K. Furthermore, the variation of the large moment agrees with a Brillouin law with effective J = 1/2, supporting the doublet ground state inferred from specific heat measurements (see section \ref{SH}).

\section {Summary and conclusions}

The Np-based intermetallic compound NpPdSn has been synthesized and characterized by x-ray diffraction, heat capacity, magnetoresistivity, thermoelectric power, Hall effect, $^{237}$Np M\"{o}ssbauer spectroscopy, and neutron scattering measurements. NpPdSn crystalizes in the ZrNiAl-type hexagonal structure and orders antiferromagnetically below 19 K and shows localized magnetism of Np$^{3+}$ ion (with doublet CEF ground state). In the magnetic state the electrical resistivity and heat capacity are dominated by electron-magnon scattering with a magnon spectrum typical of anisotropic antiferromagnets. An enhanced Sommerfeld coefficient and typical behavior of magnetorestistivity, Seebeck and Hall coefficients are all characteristic of systems with strong electronic correlations interacting with crystal electric field. The low temperature antiferromagnetic state of NpPdSn is verified by neutron diffraction and $^{237}$Np M\"{o}ssbauer spectroscopy and possible magnetic structures are discussed. Neutron diffraction and $^{237}$Np M\"{o}ssbauer spectroscopy confirm antiferromagnetic ordering in NpPdSn with Np moments confined to the $c$-axis direction. Single crystal neutron diffraction would greatly help in confirming the magnetic structure, but the difficulty of obtaining single crystal neptunium samples precludes this at present time.

\section{Acknowledgments}

We are grateful to R. Jardin and J. Rebizant for sample preparation and characterization and D. Bouexi\'{e}re and H. Thiele for technical assistance. The high-purity Np metal required for the synthesis of NpPdSn was made available in the framework of the collaboration with the Lawrence Livermore and Los Alamos National Laboratories and the US Department of Energy. Work at INL was supported by DOE's Early Career Research Program.

\end{document}